%% file: eprintFinal.tex
\documentclass[10pt]{article}
\usepackage{graphicx}
\usepackage{subfigure}
\usepackage{amsmath}

\def\Title#1{\begin{center} {\Large #1 } \end{center}}
\def\Author#1{\begin{center}{ \sc #1} \end{center}}
\def\Address#1{\begin{center}{ \it #1} \end{center}}

\newcommand\pubblock{\rightline{\begin{tabular}{l} Proceedings of the Fifth Annual LHCP\\ \pubnumber\\
         \pubdate  \end{tabular}}}

\newenvironment{Abstract}{\begin{quotation} \begin{center} 
             \large ABSTRACT \end{center}\bigskip 
      \begin{center}\begin{large}}{\end{large}\end{center} \end{quotation}}

\newenvironment{Presented}{\begin{quotation} \begin{center} 
             PRESENTED AT\end{center}\bigskip 
      \begin{center}\begin{large}}{\end{large}\end{center} \end{quotation}}

\input econfmacros.tex

\usepackage{color}

\newcommand\pubnumber{CMS-CR-2017-206 }

\newcommand\pubdate{\today}

\def\affiliation{
On behalf of the CMS Experiment, \\
Institute of High Energy Physics, Chinese Academy of Sciences\\
100049 Beijing, CHINA}

\begin{document}

\large
\begin{titlepage}
\pubblock

\vfill
\Title{Search for single production of a vector-like T quark decaying to tZ with CMS at $\sqrt{s}$ = 13 TeV  }
\vfill

\Author{Aniello Spiezia}
\Address{\affiliation}
\vfill
\begin{Abstract}
We present a search for single production of heavy vector-like quarks (VLQs), carried out by the CMS collaboration analyzing LHC pp collisions at 13 TeV. The vector like quark is a massive top quark partner that is searched for in a mass range between 0.7 and 1.7 TeV and a width between $<$1\% and 30\%. Single production can be dominant over pair production, depending on the mass of the new quark. The search is performed in a variety of final states including boosted topologies that can increase the sensitivity of the analysis.

\end{Abstract}
\vfill

\begin{Presented}
The Fifth Annual Conference\\
 on Large Hadron Collider Physics \\
Shanghai Jiao Tong University, Shanghai, China\\ 
May 15-20, 2017
\end{Presented}
\vfill
\end{titlepage}
\def\thefootnote{\fnsymbol{footnote}}
\setcounter{footnote}{0}
%

\normalsize 


\section{Introduction}
\begin{figure}[!h]
\begin{center}
\subfigure{\includegraphics[scale=0.25]{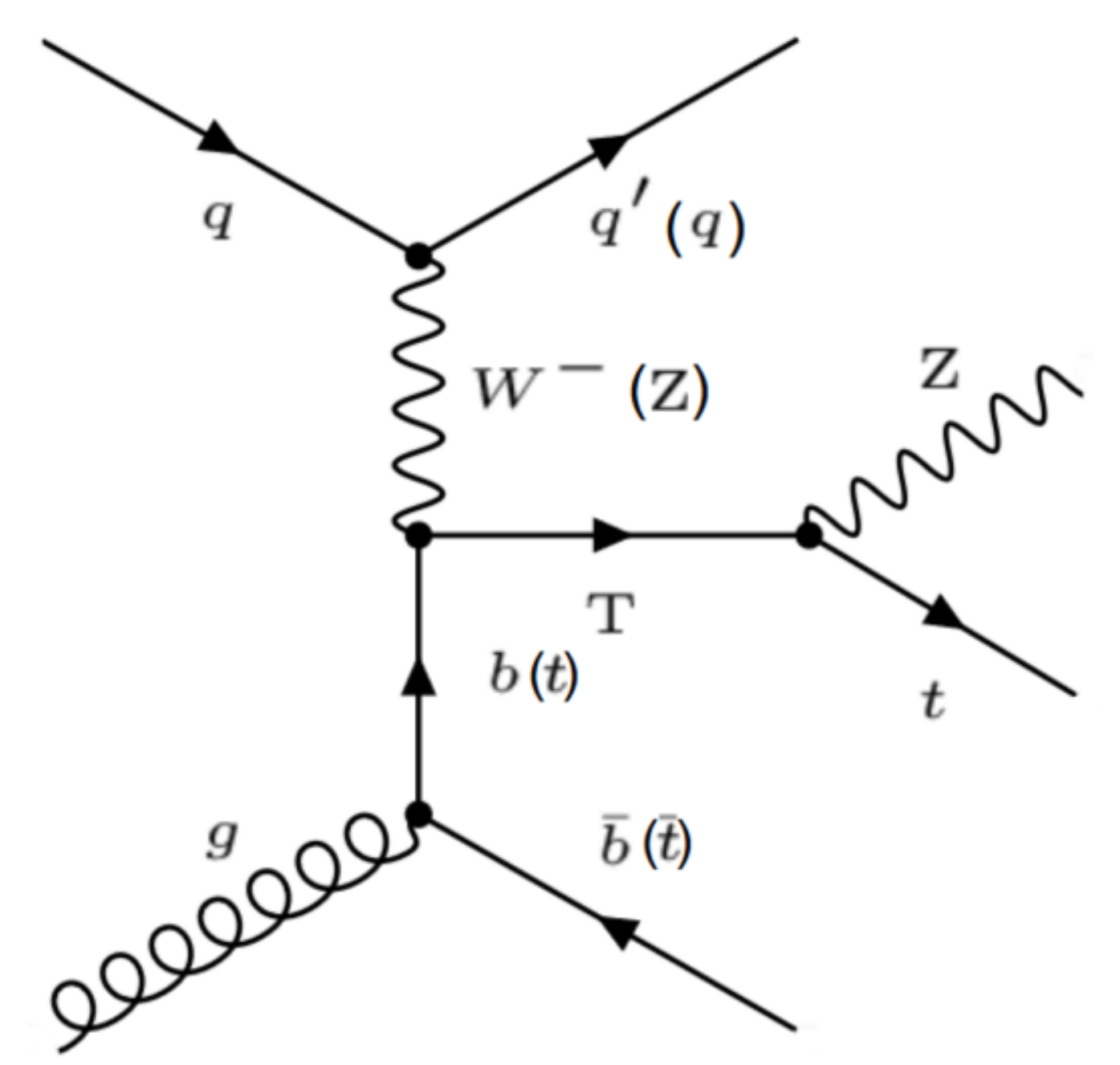}\hspace{5mm}}
\subfigure{\includegraphics[scale=0.25]{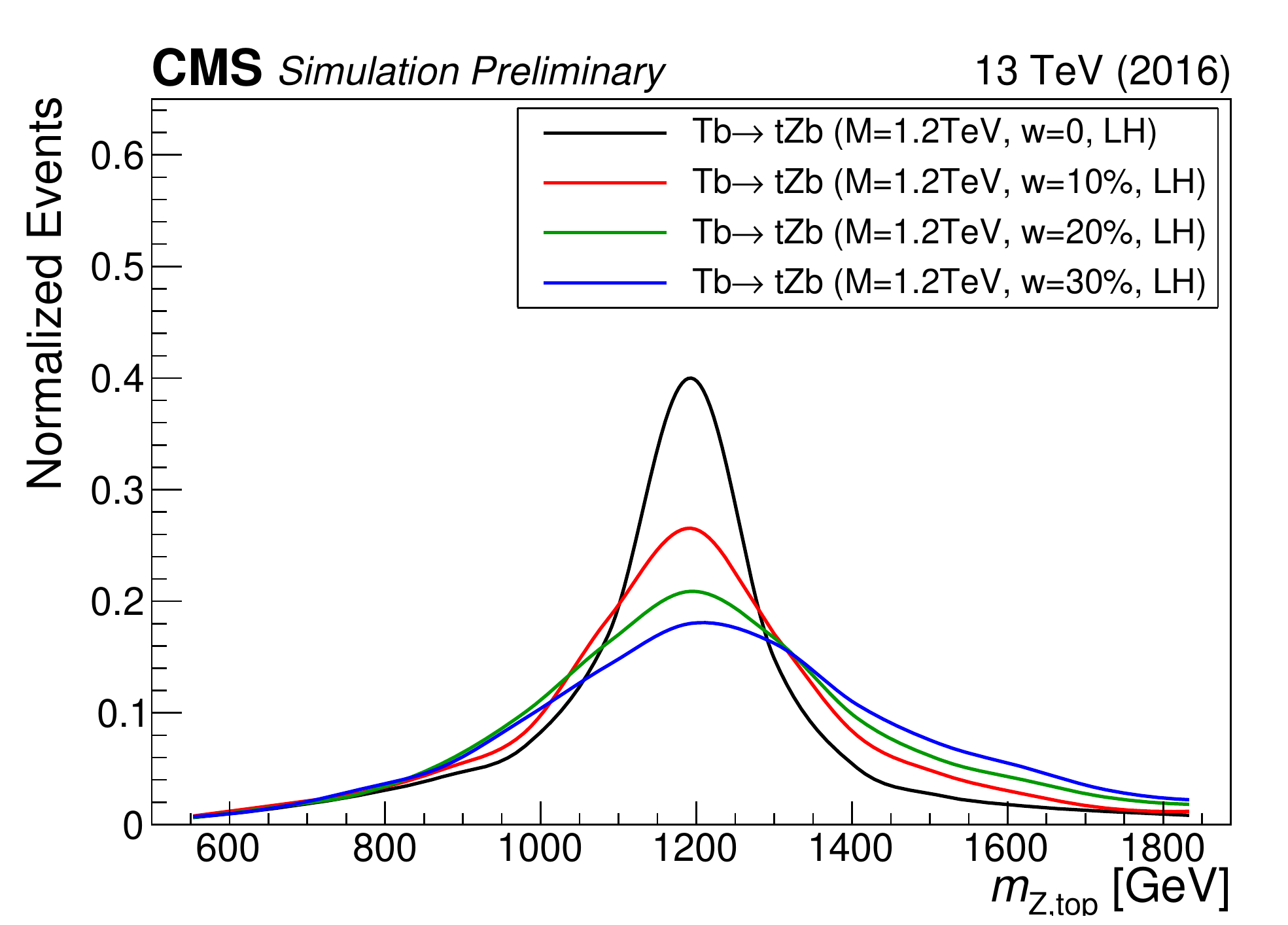}\hspace{5mm}}
\subfigure{\includegraphics[scale=0.25]{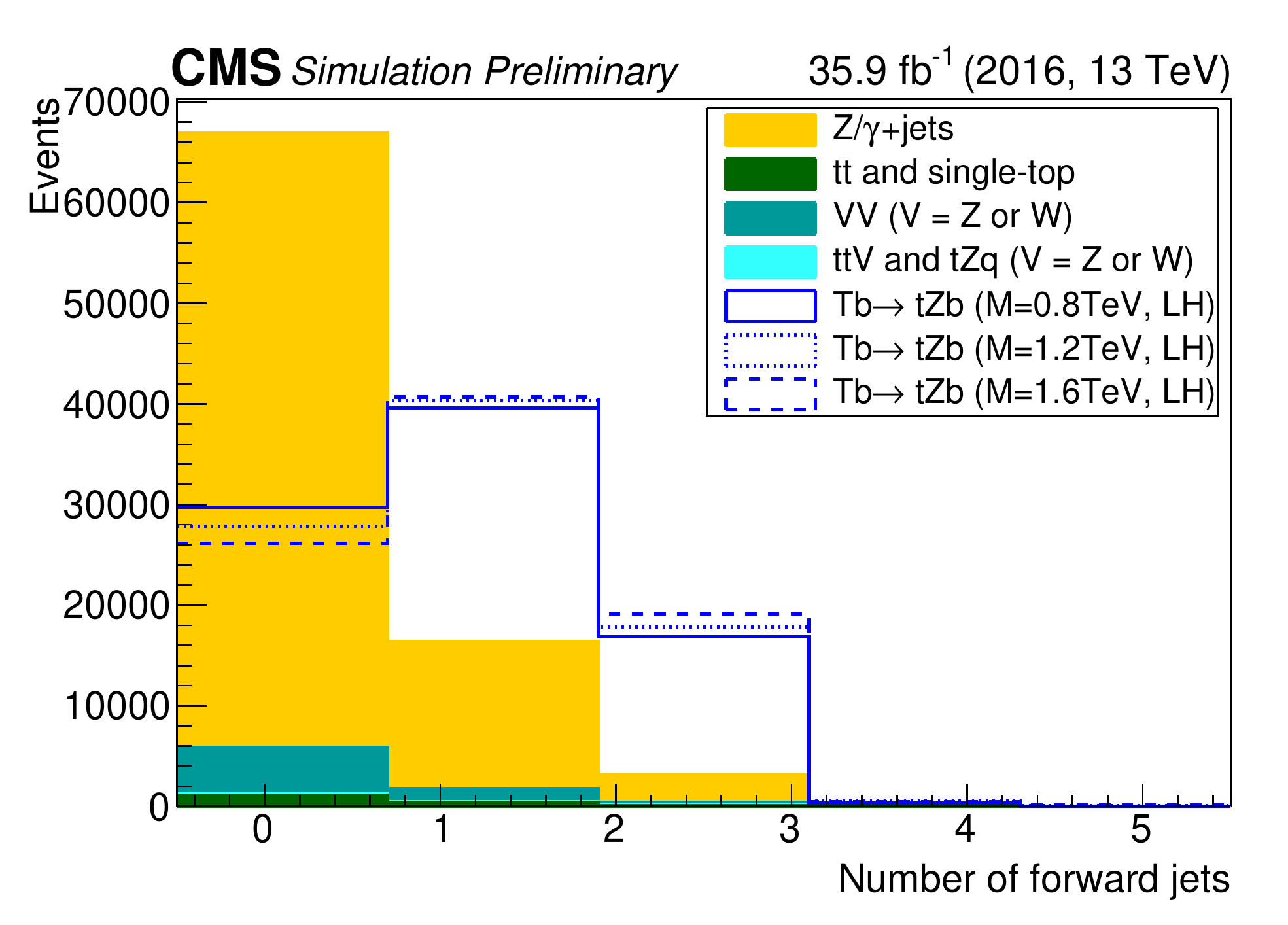}}
\caption{Leading-order Feynman diagram for the single production of a T quark, either in association with a b quark or a t quark (left). Reconstructed T mass for the T(b) signal, with T mass equal to 1.2 TeV and four values of the T width: negligible, 10\%, 20\% and 30\% (center). Comparison between background and signal for the number of forward jets after event preselection (right). Figures taken from~\cite{Sirunyan:2017ynj}.}\label{fig:feynman}
\end{center}
\end{figure}

\vspace{-1mm}
Vector-like quarks (VLQs)~\cite{Aguilar-Saavedra:2013qpa, AguilarSaavedra:2009es, DeSimone:2012fs, Matsedonskyi:2014mna, Buchkremer:2013bha} are hypothetical new particles introduced to address some of the problems related to the nature of electroweak symmetry breaking. We study the single production of a vector-like T quark with charge $+2/3$ in its decay to a Z boson and a t quark~\cite{Sirunyan:2017ynj}: the final state consists of the Z boson decaying to electrons or muons and the t quark decaying to quarks, i.e. t $\rightarrow$ Wb $\rightarrow$ q$^{\prime}$ $\bar{\mbox{q}}$b. The search is performed using the data collected by the CMS experiment~\cite{Chatrchyan:2008aa} at 13 TeV. In Fig.~\ref{fig:feynman} (left), one of the LO Feynman diagram for the single production of a T quark is shown. From the diagram it is clear that the T quark is produced in association with either a b quark, denoted T(b), or a t quark, denoted T(t), and with an additional quark produced in the forward region of the detector. The T quark has three decay channels into SM particles: bW, tZ, and tH. The equivalence theorem implies that the branching fractions for the T quark into tZ is 0.25, in case the T is a singlet of the SM, or 0.5, it is a doublet. The T quark can have both left-handed (LH) and right-handed (RH) couplings to SM particles. Several T quark width hypotheses are studied: from negligibly small to larger widths (10, 20, and 30\% of the T quark mass). The reconstructed T mass for the four different width hypotheses are shown in Fig.~\ref{fig:feynman} (center).

\begin{table}[!b]
\centering
\caption{Summary of the final event selection.}\label{table:EventSelection}
\begin{tabular}{lcccccc} \hline
                        & \scriptsize 2$\mu$+1t jet & \scriptsize 2e+1t jet & \scriptsize 2$\mu$+1W jet+1b jet & \scriptsize 2e+1W jet+1b jet & \scriptsize 2$\mu$+1b jet+2 jets & \scriptsize 2e+1b jet+2 jets\\\hline
Leptons                 & 2 muons         & 2 electrons & 2 muons                  & 2 electrons          & 2 muons                  & 2 electrons         \\[0.5ex]
Lead lep $\mbox{p}_{\mbox{T}}$           & $>$120 GeV& $>$120 GeV& $>$120 GeV& $>$120 GeV& $>$120 GeV& $>$120 GeV \\[0.5ex]
$\Delta R(\ell,\ell)$   & {$<$1.4} & {$<$1.4}       & {$<$0.6} & {$<$0.6} & {$<$0.6} & {$<$0.6}\\[0.5ex]
Jet & \multicolumn{2}{c}{1 top jet} & \multicolumn{2}{c}{1 W jet, 1 b jet} & \multicolumn{2}{c}{3 AK4 jets (one b-tagged)} \\[0.5ex]
top $\mbox{p}_{\mbox{T}}$               & $>$400 GeV     & $>$400 GeV     & {$>$150 GeV}   & {$>$150 GeV}    & {$>$150 GeV}    & {$>$150 GeV}    \\[0.5ex]
N(b jet)                & {$\geq$1} & {$\geq$1} & {$\geq$1} & {$\geq$1} & {$\geq$1} & {$\geq$1} \\\hline
\end{tabular}
\end{table}

\section{Event selection}\vspace{-1mm}
The presence of a Z boson and of additional jets from the t quark decay results in a background largely dominated by Z/$\gamma$*+jets events ($>$ 80\%). Smaller contributions from other sources are still present (t$\bar{\mbox{t}}$+V, tZq, t$\bar{\mbox{t}}$, single t quark, and SM VV diboson production, where V represents a W or Z boson). The event selection requires the selection of events with two oppositely charged leptons (either muons or electrons) forming a Z boson with an invariant mass between 70 and 110 GeV. The t quark from the T quark decay is identified in three different ways: fully merged (a t jet is identified), partially merged (a W jet and a b jet are identified), or resolved (three AK4 jets are reconstructed). The experimental strategy consists in defining ten event categories, depending on how the Z boson or the t quark candidates are reconstructed and on the number of forward jets present in the event. The presence of a forward jet is an interesting feature of the signal under study and a comparison between background and signal for the number of forward jets is shown in Fig.~\ref{fig:feynman} (right). In addition to requiring a Z boson and a t quark in the event, other requirements are optimized to increase the sensitivity of the analysis. The final event selection is summarized in Table~\ref{table:EventSelection}.

\section{Background estimate}
\begin{figure}[!h]
\begin{center}
\subfigure{\includegraphics[scale=0.25]{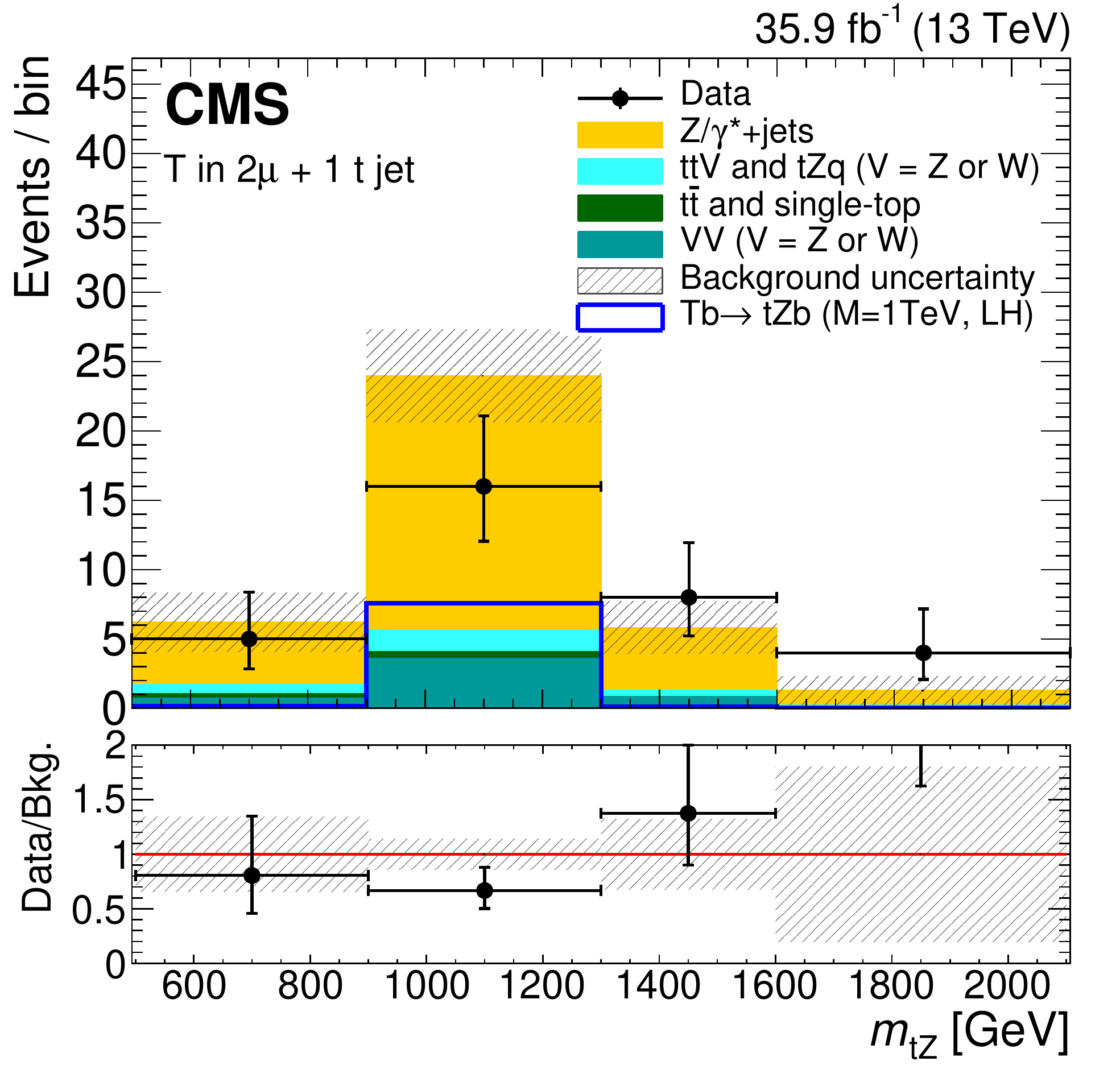}\hspace{5mm}}
\subfigure{\includegraphics[scale=0.25]{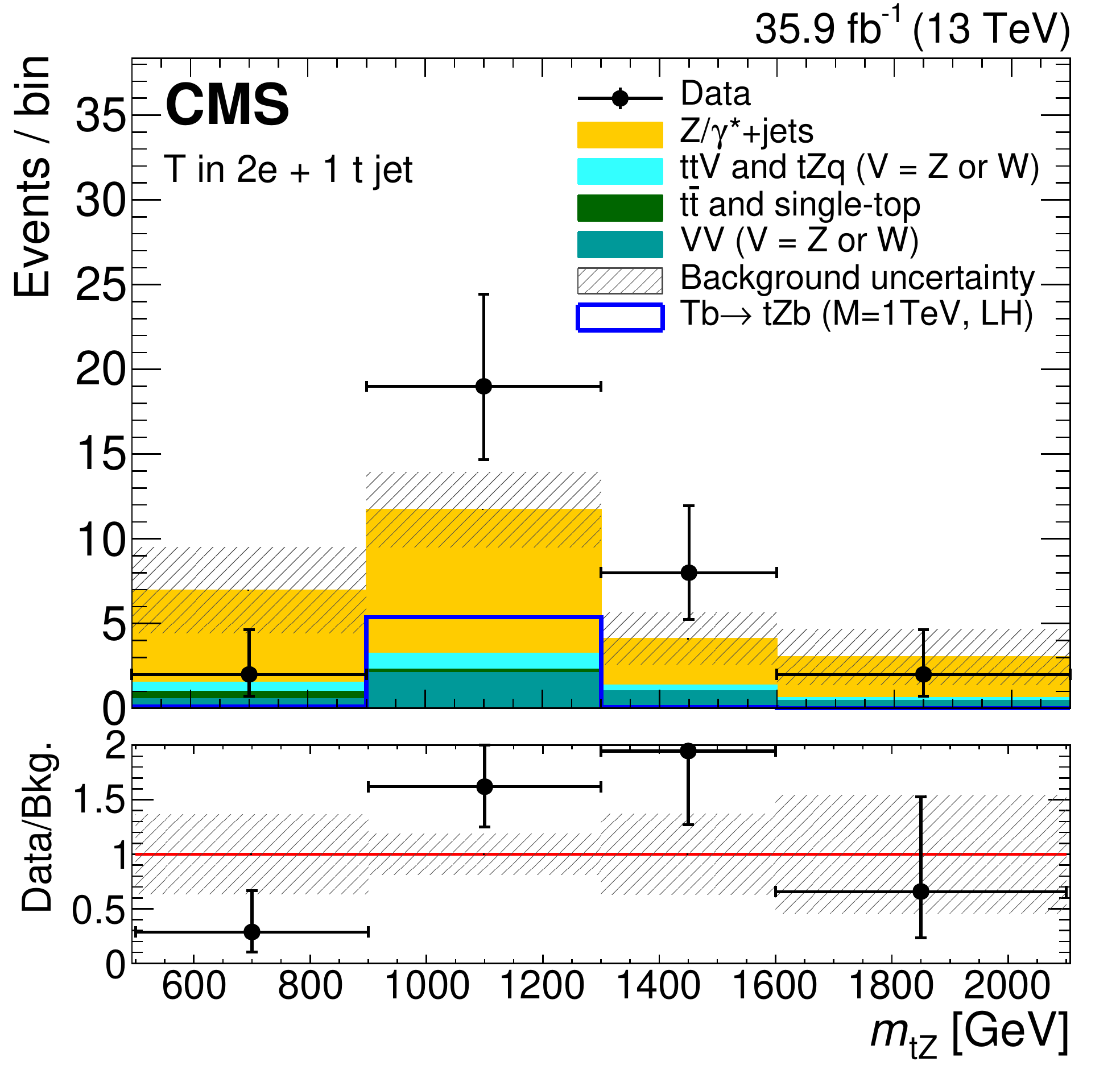}\hspace{5mm}}
\subfigure{\includegraphics[scale=0.25]{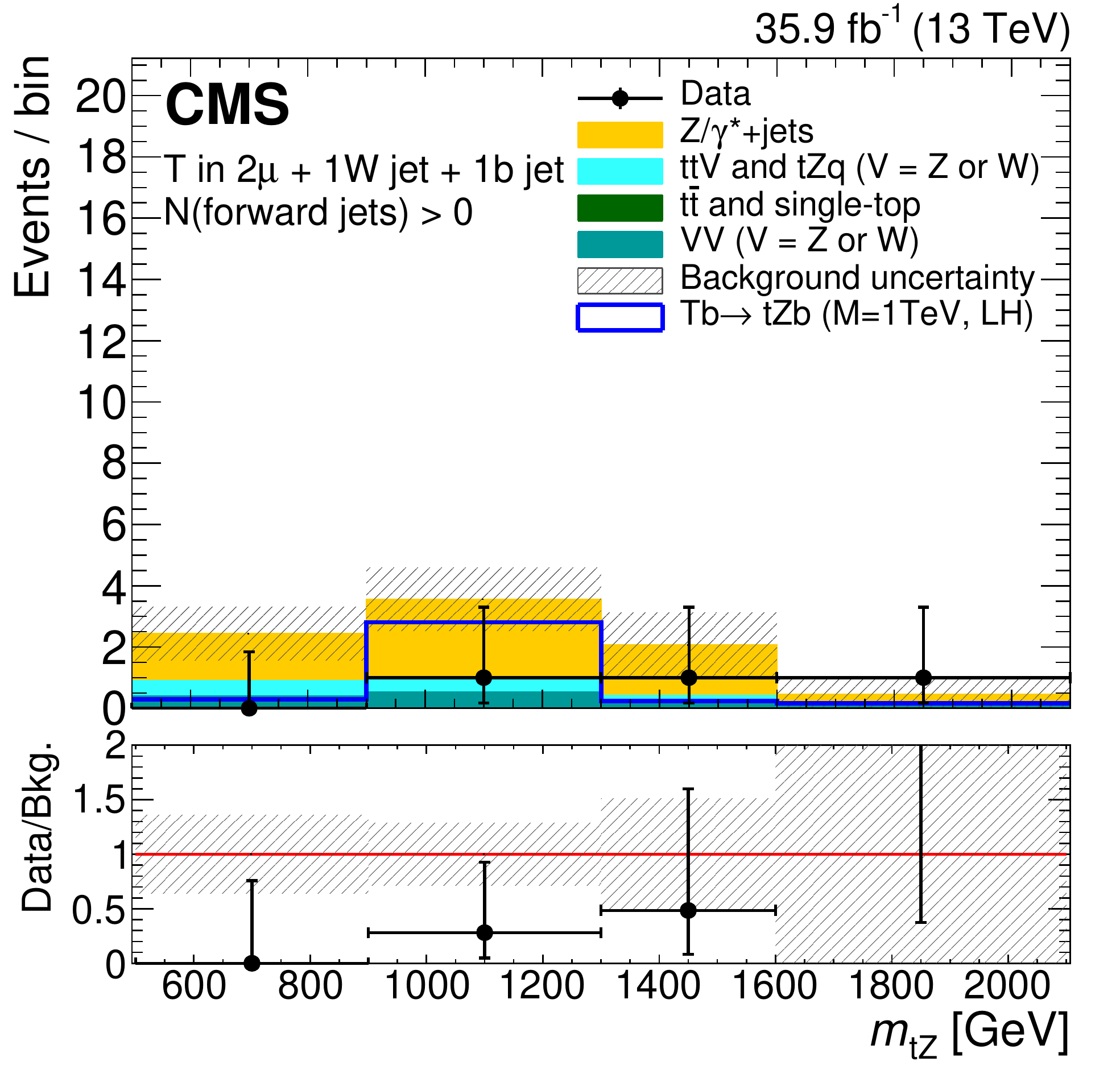}}\\\vspace{-3mm}
\subfigure{\includegraphics[scale=0.25]{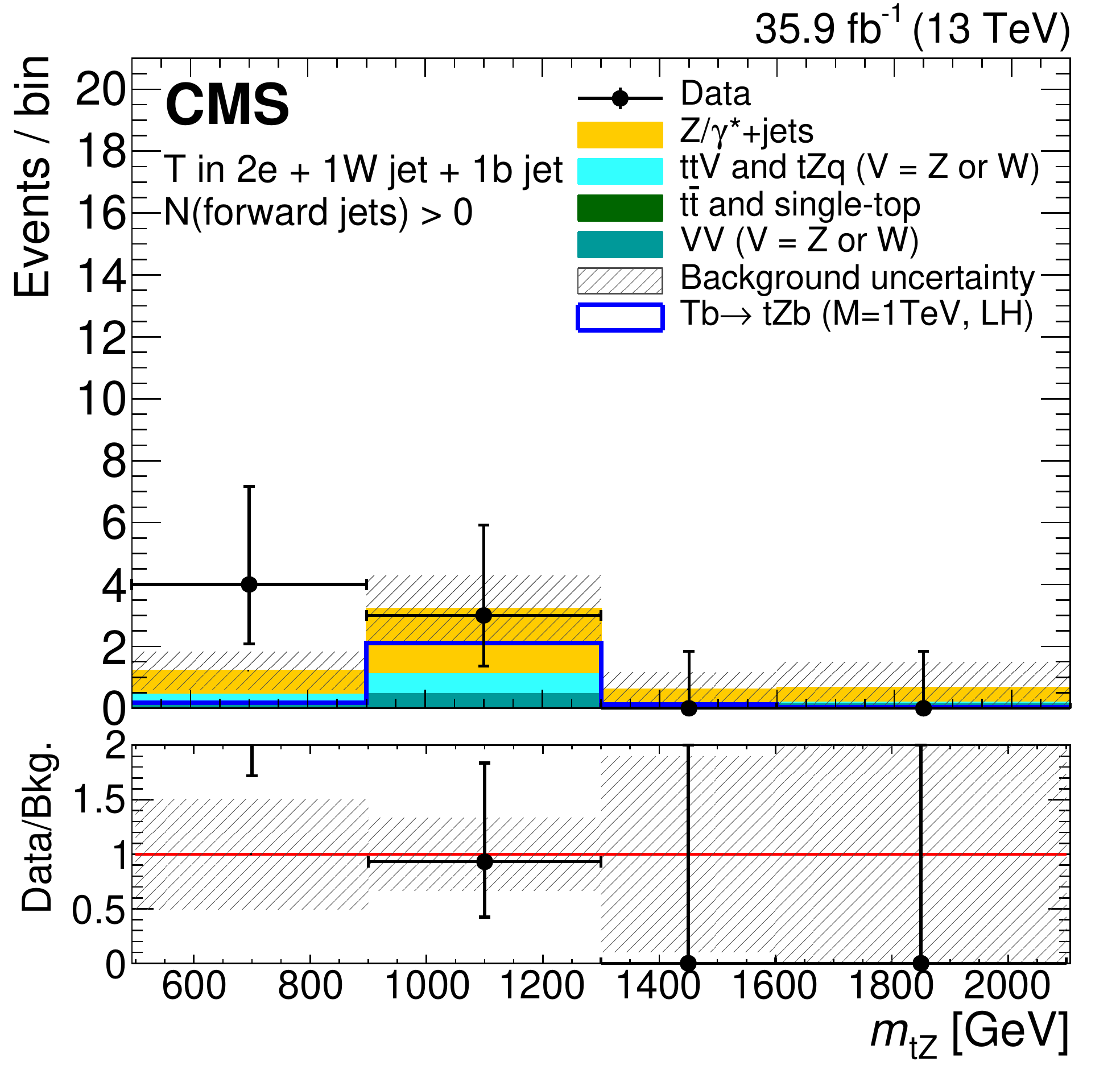}\hspace{5mm}}
\subfigure{\includegraphics[scale=0.25]{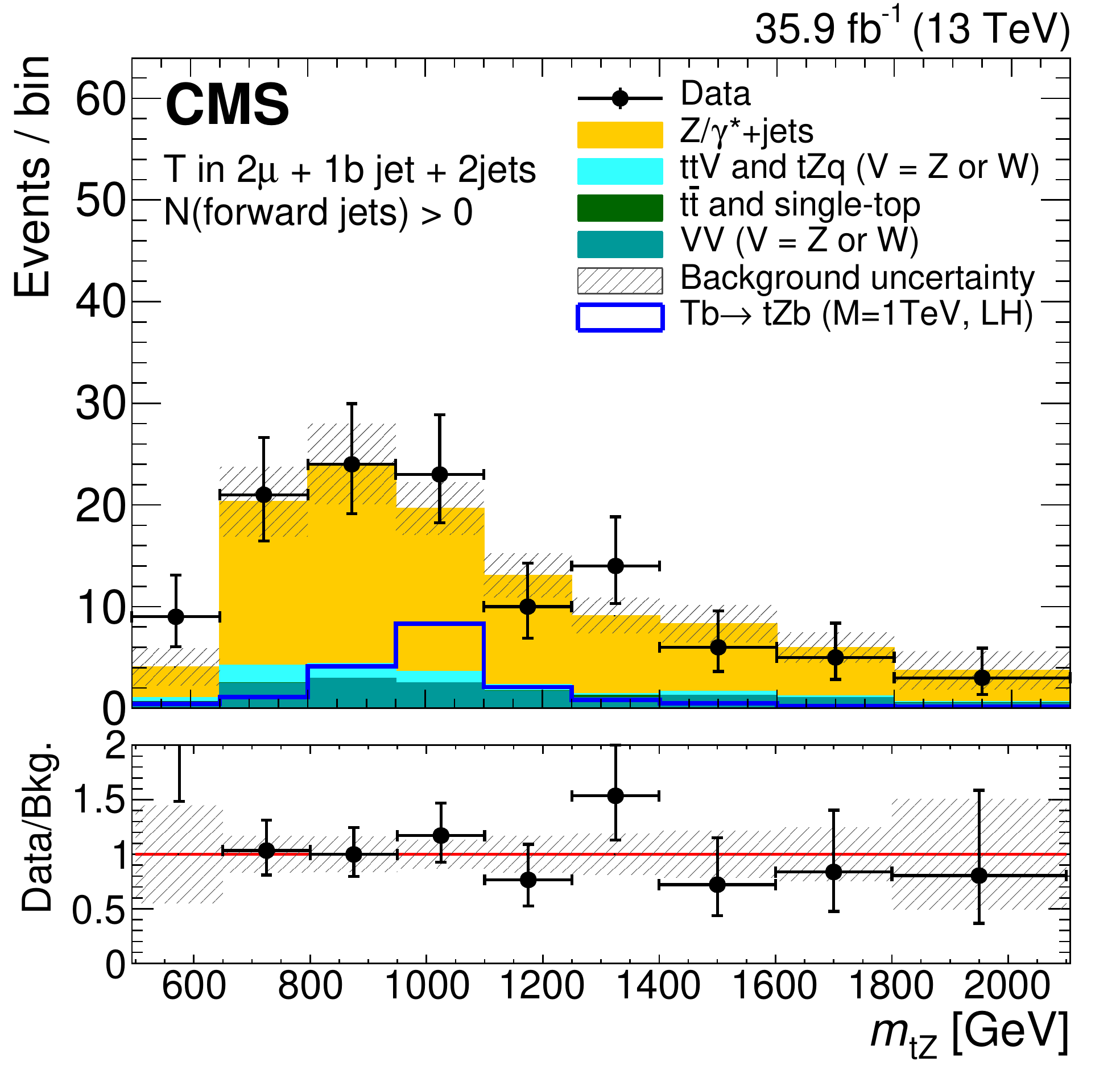}\hspace{5mm}}
\subfigure{\includegraphics[scale=0.25]{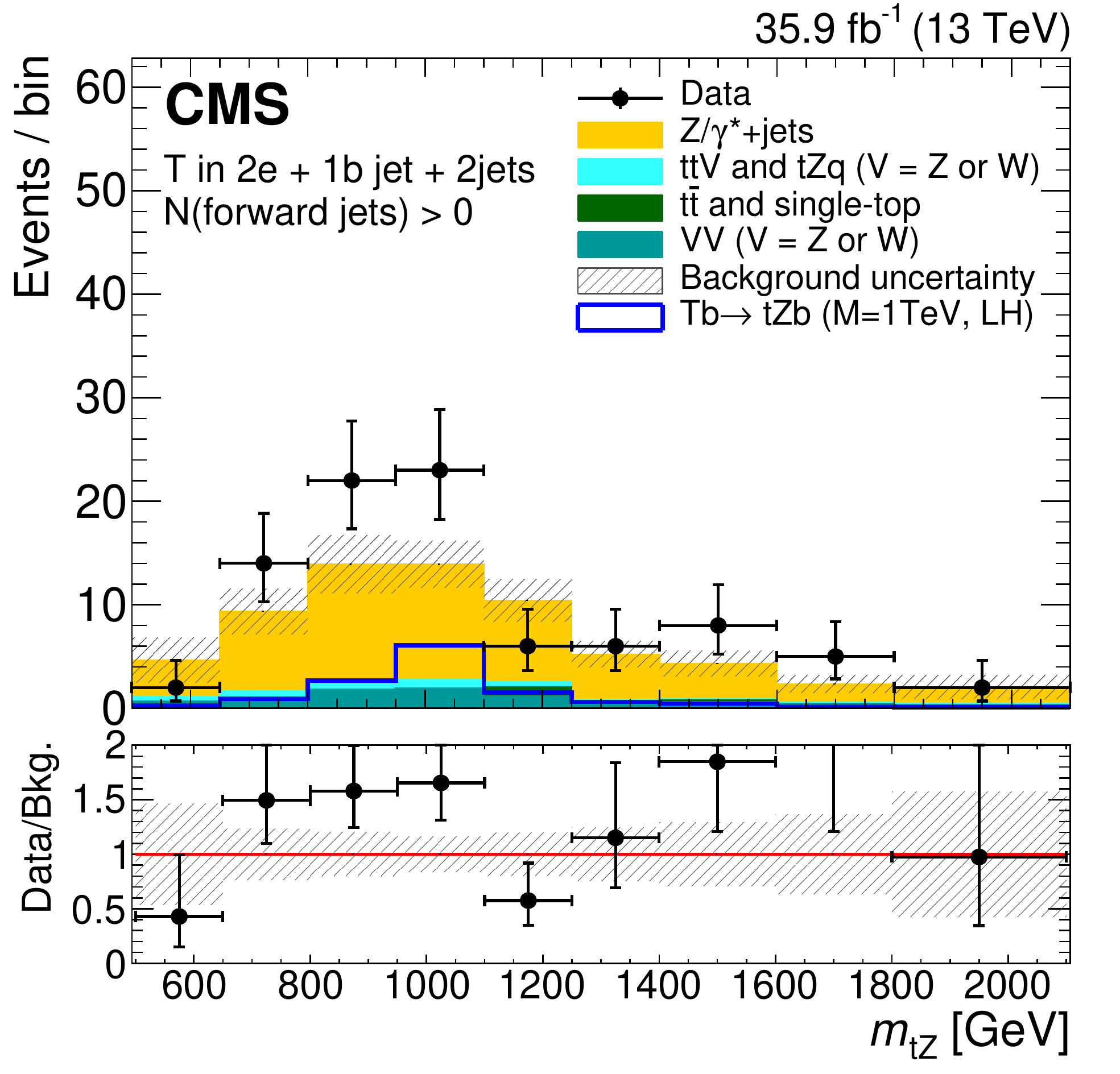}}
\caption{Comparison between the data, the background estimate, and the expected signal for the 6 categories with the highest sensitivity. The background composition is taken from simulation. Figures taken from~\cite{Sirunyan:2017ynj}.}\label{fig:background}
\end{center}
\end{figure}

\vspace{-1mm}
The background estimate is primarily based on a control sample in data through the following formula:\vspace{-1mm}
$$N_{\text{bkg}}(\text{M}_{\text{t,Z}}) = N_{\text{cr}}(\text{M}_{\text{t,Z}}) \cdot \alpha(\text{M}_{\text{t,Z}}),$$
\begin{itemize}\vspace{-2.5mm}
\item $N_{\text{cr}}(\text{M}_{\text{t,Z}})$: number of events in the data sample in the control region\vspace{-1mm}
\item $\alpha(\text{M}_{\text{t,Z}})$: ratio of  number of events in signal region to the ones in control region, from simulation\vspace{-1mm}
\end{itemize}
The control region from which the number of events is extrapolated into the signal region is defined by the full event selection, but applying a veto on the presence b tagged jets. A comparison between data and background estimate for the six categories with the highest sensitivity is shown in Fig.~\ref{fig:background}.

\section{Systematic Uncertainties}\vspace{-1mm}
The systematic uncertainties for the signal come from corrections that are applied to the simulation in order to match distributions in data. The uncertainty in the background estimate comes from five sources: statistical uncertainties of data and simulation in the control region, the small differences between observation and prediction for a closure test, the uncertainty related to possible mismodelling of the Z+light quark and Z+b quark fractions in simulation, and the uncertainty from the b tagging efficiency differences in data and simulation.

\section{Results}
\begin{figure}[!h]
\begin{center}\vspace{-1mm}
\subfigure{\includegraphics[scale=0.2]{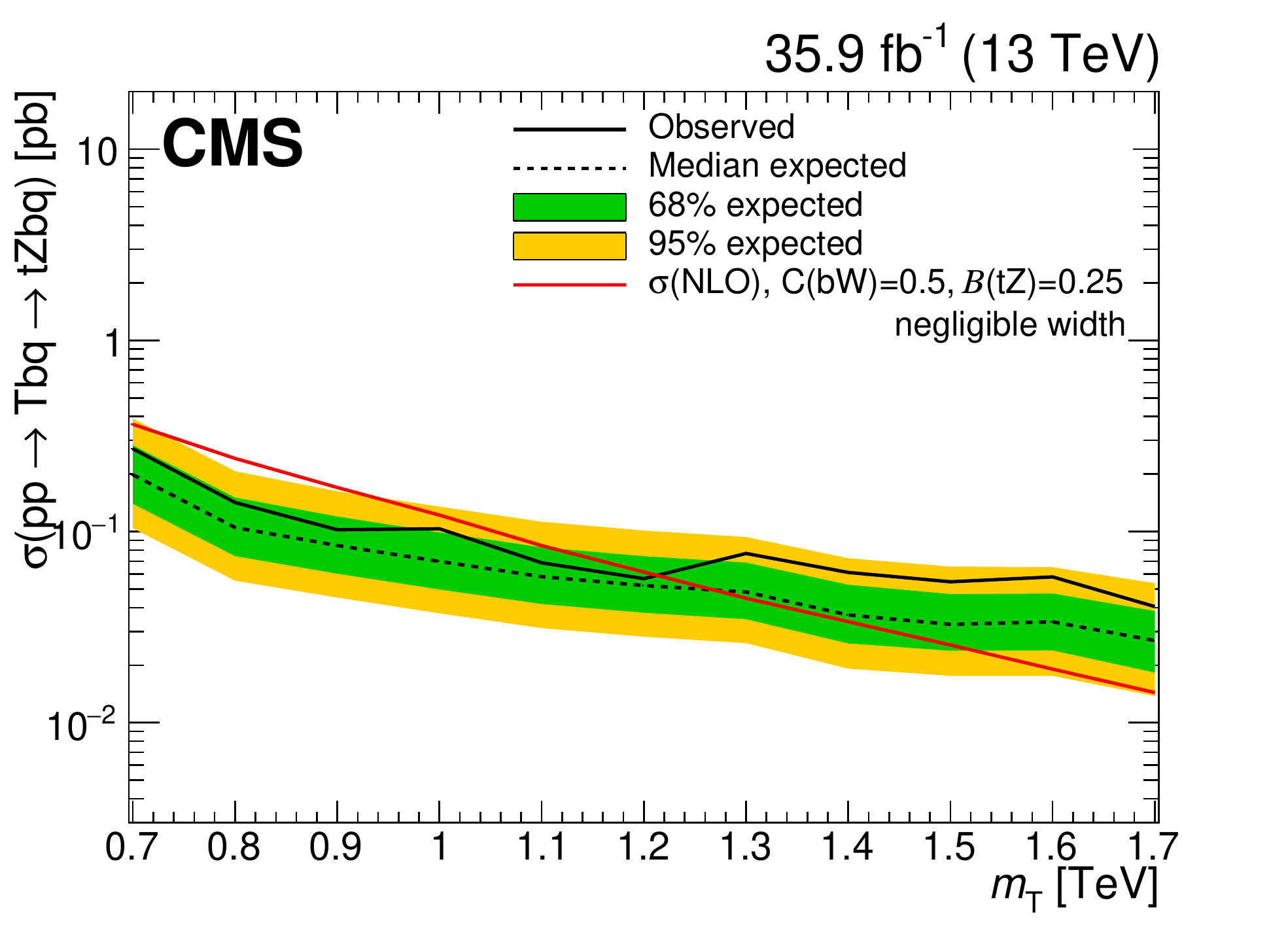}}
\subfigure{\includegraphics[scale=0.2]{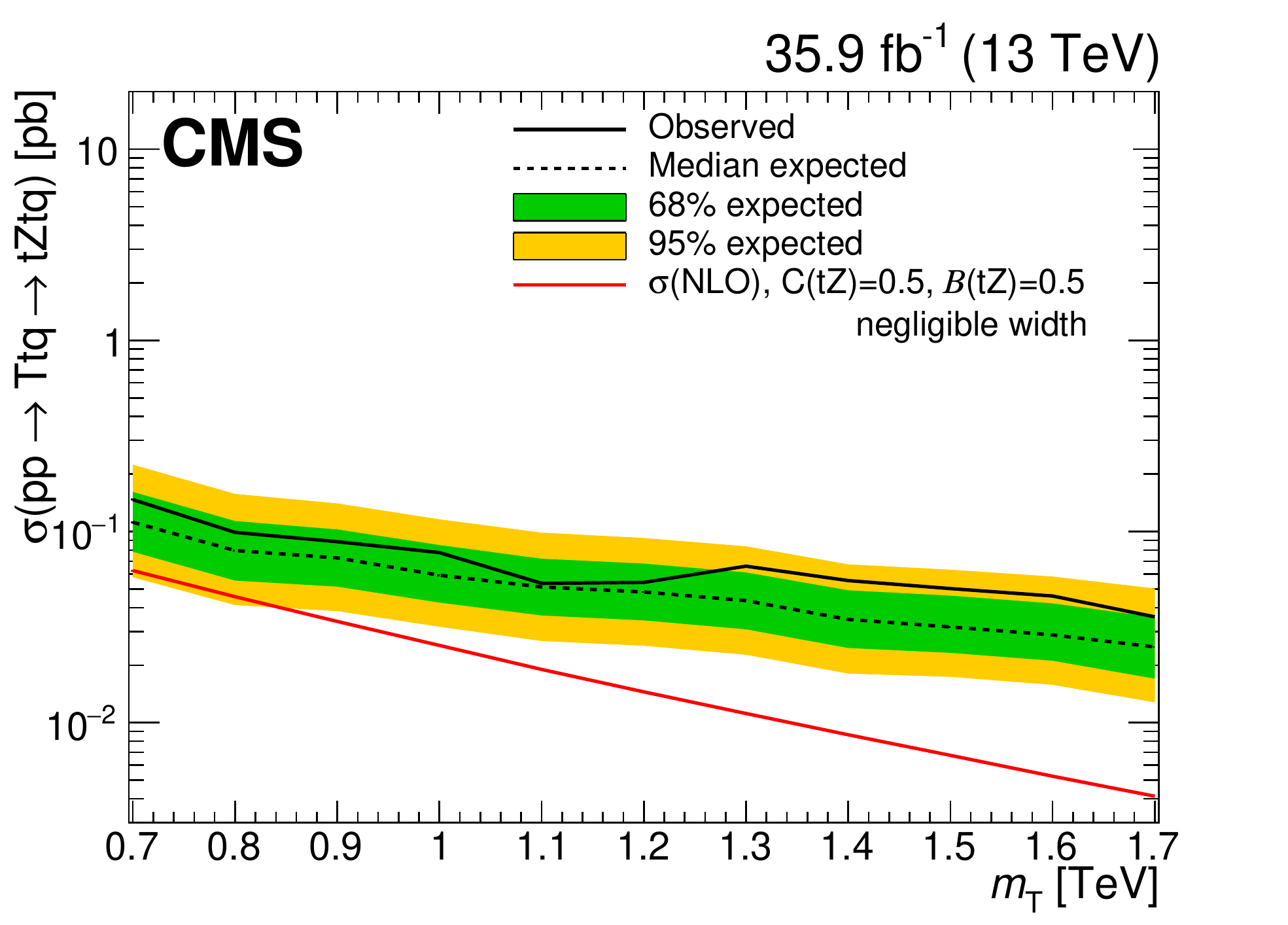}}
\subfigure{\includegraphics[scale=0.2]{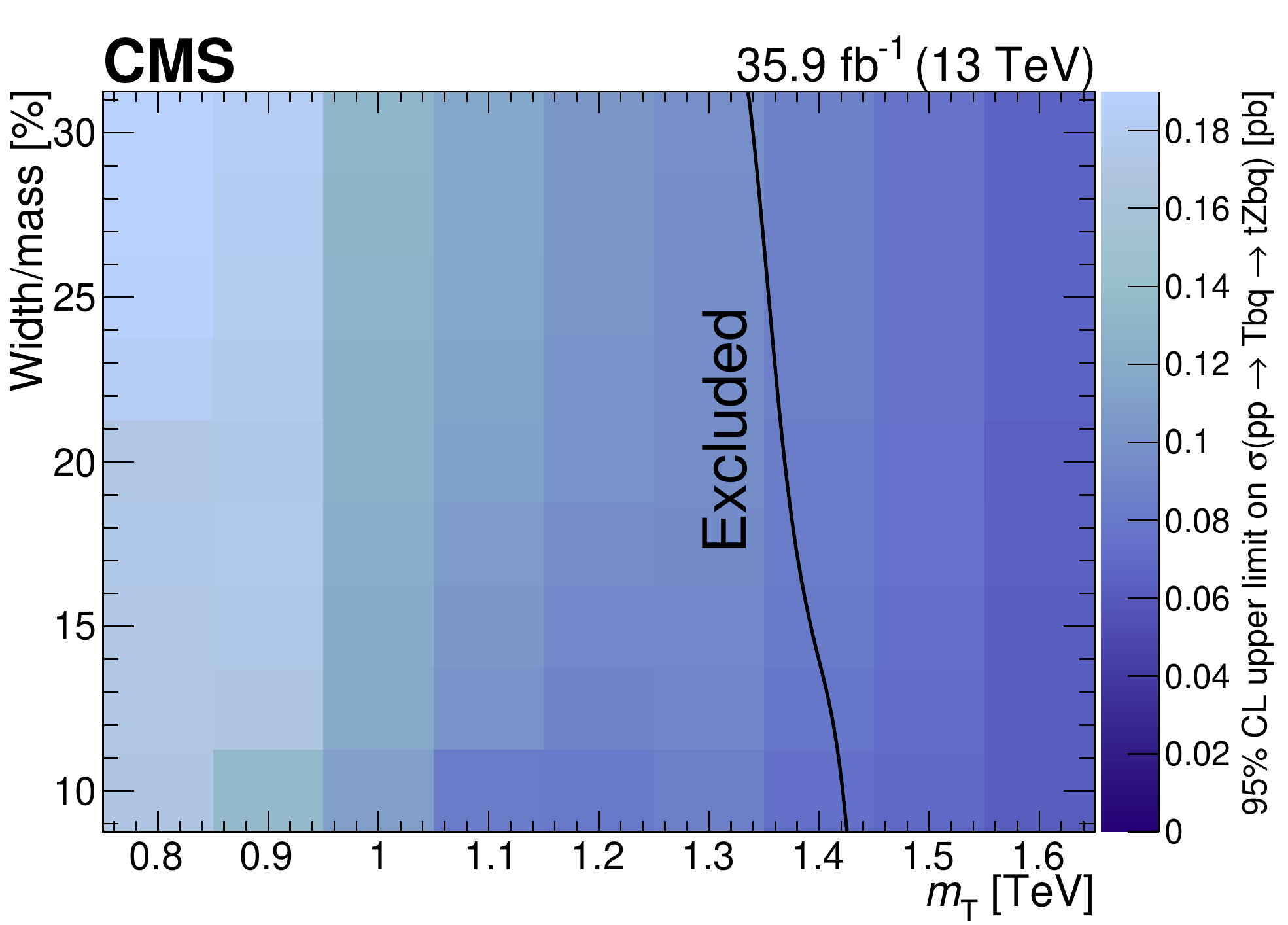}}
\subfigure{\includegraphics[scale=0.2]{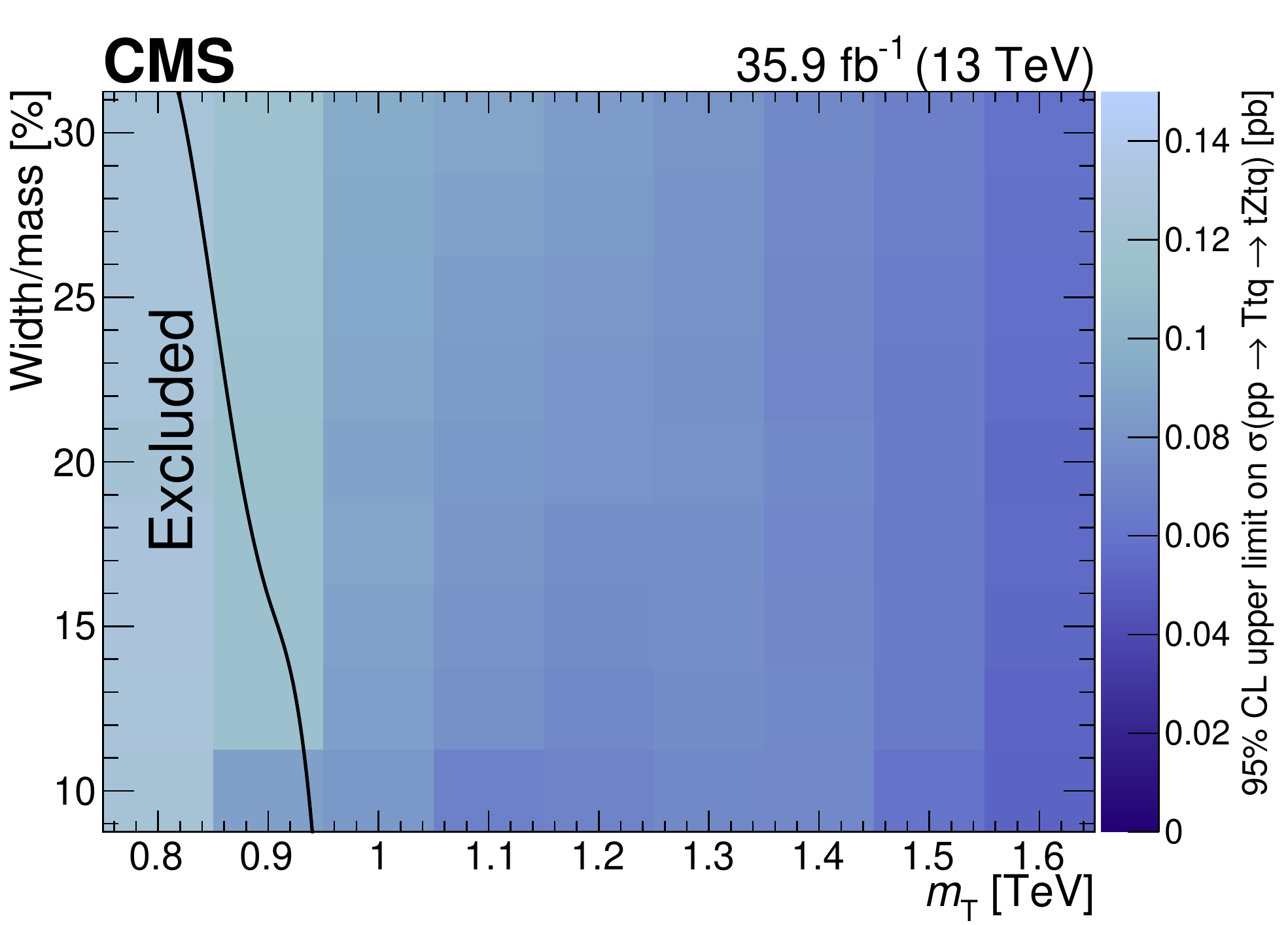}}
\caption{Observed and expected limits at 95\% CL on the product of cross section and branching fraction for singlet LH T(b) (left) and doublet RH T(t) (center-left) signals, for a T quark with narrow width. Observed limits for larger T width hypotheses for T(b) (center-right) and T(t) (right) signals. Figures from~\cite{Sirunyan:2017ynj}.}\label{fig:results}\vspace{-5mm}
\end{center}
\end{figure}

\vspace{-1mm}
As shown in Fig.~\ref{fig:background}, no significant deviations from the expected background are observed. We set upper limits on the product of the cross section and branching fraction of a T quark decaying to tZ. In Fig.~\ref{fig:results}, the observed and expected limits are shown in the negligibly small width, for the singlet LH T(b) (left) and doublet RH T(t) (center-left) production modes. Upper limits are compared with theoretical cross sections calculated at NLO in Ref.~\cite{Matsedonskyi:2014mna}. For this model, a singlet LH T quark is excluded at 95\%~CL for masses below 1.2~TeV. In the same Fig.~\ref{fig:results}, the observed upper limits at 95\% CL are shown as a function of the T quark width and T quark mass in the ranges from 10 to 30\% and 0.8 to 1.6 TeV, respectively. A sensitivity similar to that obtained assuming a narrow-width T quark is observed. In this case the experimental results are compared with the theoretical cross sections calculated at LO using a modified version of the model constructed by the authors of~\cite{Buchkremer:2013bha,Fuks:2016ftf,Oliveira:2014kla}. For this model, the data exclude a singlet LH T quark produced in association with a b quark, for masses below values in the range 1.34 and 1.42~TeV depending on the width and a doublet RH T quark produced in association with a t quark is excluded for masses below values in the range 0.82 and 0.94~TeV.


\end{document}

%% file: econfmacros.tex
\def\beq{\begin{equation}}
\def\eeq#1{\label{#1}\end{equation}}
\def\eeqn{\end{equation}}

\def\beqa{\begin{eqnarray}}
\def\eeqa#1{\label{#1}\end{eqnarray}}
\def\eeqan{\end{eqnarray}}

\let\bar=\overbar

\def\Dslash{\not{\hbox{\kern-4pt $D$}}}
\def\dslash{\not{\hbox{\kern-2pt $\del$}}}

\def\msb{{\bar{\ssstyle M \kern -1pt S}}}




%% file: eprintFinal.bbl
\begin{thebibliography}{99}\vspace{-1mm}
\bibitem{Aguilar-Saavedra:2013qpa} 
  J.~A.~Aguilar-Saavedra, R.~Benbrik, S.~Heinemeyer and M.~Pérez-Victoria,
  Phys.\ Rev.\ D {\bf 88}, no. 9, 094010 (2013)
  arXiv:1306.0572 [hep-ph].
  
\bibitem{AguilarSaavedra:2009es} 
  J.~A.~Aguilar-Saavedra,
  JHEP {\bf 0911}, 030 (2009)
  arXiv:0907.3155 [hep-ph].
  
\bibitem{DeSimone:2012fs} 
  A.~De Simone, O.~Matsedonskyi, R.~Rattazzi and A.~Wulzer,
  JHEP {\bf 1304}, 004 (2013)
  arXiv:1211.5663 [hep-ph].
  
\bibitem{Matsedonskyi:2014mna} 
  O.~Matsedonskyi, G.~Panico and A.~Wulzer,
  JHEP {\bf 1412}, 097 (2014)
  arXiv:1409.0100 [hep-ph].
  
\bibitem{Buchkremer:2013bha} 
  M.~Buchkremer, G.~Cacciapaglia, A.~Deandrea and L.~Panizzi,
  Nucl.\ Phys.\ B {\bf 876}, 376 (2013)
  arXiv:1305.4172 [hep-ph].
  
\bibitem{Sirunyan:2017ynj} 
  A.~M.~Sirunyan {\it et al.} [CMS Collaboration],
  arXiv:1708.01062 [hep-ex].
  
\bibitem{Chatrchyan:2008aa} 
  S.~Chatrchyan {\it et al.} [CMS Collaboration],
  JINST {\bf 3}, S08004 (2008).
  
\bibitem{Fuks:2016ftf} 
  B.~Fuks and H.~S.~Shao,
  Eur.\ Phys.\ J.\ C {\bf 77}, no. 2, 135 (2017)
  arXiv:1610.04622 [hep-ph].
  
\bibitem{Oliveira:2014kla} 
  A.~Oliveira,
  arXiv:1404.0102 [hep-ph].

\end{thebibliography}
